\documentclass[final,usletter,number,sort&compress]{article}

\usepackage[top=1.in, bottom=1.in, left = 1in, right=1in]{geometry}

\usepackage{graphicx}
\usepackage{dcolumn}
\usepackage{bbm}
\usepackage{textcomp}
\usepackage{color}
\usepackage{gensymb}
\usepackage{amsmath}
\usepackage{wasysym}

\begin{document}

\def\nuc#1#2{${}^{#1}$#2}

\title{Characterization of protonated and deuterated Tetra-Phenyl Butadiene Film in a Polystyrene Matrix}
\author{
V.M. Gehman\footnote{Corresponding author: vmgehman@lbl.gov}\\Lawrence Berkeley National Laboratory, Berkeley, CA 94720
\and
T.M. Ito, W.C. Griffith\\Los Alamos National Laboratory, Los Alamos, NM 87545
\and
S.R. Seibert\\University of Pennsylvania, Philadelphia, PA 19104
}
\date{\today}


\maketitle

\begin{abstract}
We study the effect of deuteration and annealing on the fluorescence spectrum shape and VUV to visible conversion efficiency of TPB films in a polystyrene matrix with input light from 120 to 220 nm.  We observed no discernible difference in the fluorescence spectrum shape between any of the films.  The deuterated film performed equally well compared to the standard one in terms of conversion efficiency, but annealing seems to degrade this efficiency to roughly 75\% of its non annealed value at all wavelengths studied.
\end{abstract}

\section{Introduction and Motivation}\label{sec:Intro}
Various experiments in search of answers to some of the most intriguing questions in physics today have been proposed, are under development, or are being conducted using liquid noble gases as particle detectors or as media in which the experiment is conducted. These include: detectors for the detection of dark matter in the form of Weakly Interacting Massive Particles \cite{XENON, XMASS, LUX, MiniCLEAN, DEAP3600, CLEAN, DARKSIDE}, neutrino and other particle detectors \cite{ICARUS, MicroBooNE, MEG, EXO, NEXT,   LBNE, Lanou1987, McKinsey2000}, searches for the neutron's electric dipole moment \cite{nEDM1, nEDM2}, and measurements of the neutron lifetime\cite{Huffman2000}.  Many of these radiation detector systems use the scintillation light from particle interactions in the noble gas.  Noble gasses have remarkably high scintillation yield (typically between 20 and 40 photons per keV\cite{Lippincott2010}).  The main complication with detecting noble gas scintillation photons is that they are well into the vacuum ultraviolet (VUV), ranging from approximately 80 nm (for helium and neon) to 128 nm (for argon) and 175 nm (for xenon).  Photons at this wavelength are quite difficult to detect because they are at a short enough wavelength to scatter off of most chemical bonds (and are therefore strongly absorbed by most materials) but are not energetic enough to be detected with some calorimetric system (as one would with x rays or $\gamma$ rays).

Since xenon scintillation photons have the longest wavelength of the noble gases commonly used as radiation detectors, some xenon-based detectors are designed to be directly sensitive to 175-nm light (these are typically those using avalanche photodiodes or phototubes with quartz windows).  Detectors using the other noble gasses require that the scintillation photons be shifted into visible light with some fluorescent material (note that depending on the details of their photon detection system, some xenon detectors also use this approach to boost their detection efficiency).  Because noble gasses are transparent to their scintillation photons, these fluorescent materials do not have to be dissolved or suspended in the bulk detector.  Instead they tend to be films deposited onto surfaces with which the noble gas is in contact.  These fluorescent films tend to be hydrocarbon plastics containing several phenyl groups.  One popular and therefore well-studied wavelength shifting (WLS) plastic is ``1,1,4,4-Tetraphenyl Butadiene'' (TPB)\cite{McKinsey1997, Jones2012, Gehman2011}. In particular, for liquid helium scintillation detection, TPB has been used in several experiments including Refs.~\cite{Huffman2000,McKinsey2004,Archibald2006,Ito2012}.

For experiments in which ultracold neutrons (UCN)\cite{Golub1991, Ignatovich1990} are stored in a volume filled with liquid helium and the walls of the UCN storage volume are coated with WLS plastic, it is important that the WLS film be made of deuterated material: regular ({\it i.e.} protonated) WLS would cause neutrons to be captured on hydrogen atoms thereby providing no neutron storage. (The use of deuterated polystyrene for UCN storage was first proposed and demonstarted by Lamoreaux~\cite{Lamoreaux1988}.) The Neutron EDM experiment \cite{nEDM1, nEDM2}, currently under development to be mounted at the Spallation Neutron Source at the Oak Ridge National Laboratory, is one such experiment. Furthermore the walls of the UCN storage cell (nominally made of PMMA) need to be annealed before and/or after the construction of the cell to properly remove the stress stored in the material for cryogenic use. Both the deuteration and annealing of the WLS films have the possibility of altering the visible fluorescence spectrum and/or the VUV-to-visible conversion efficiency from those measured in previous work.

\section{Methods and Measurements}\label{sec:Methods}
For the work presented in this article, we measured the visible fluorescence spectra and absolute fluorescence efficiency for three wavelength shifting films:
\begin{enumerate}
\item 70\% polystyrene (PS) plus 30\% tetra-phenyl butadiene\label{lis:PSplusTPB},
\item 70\% deuterated PS-d$_8$ plus 30\% deuterated tetra-phenyl butadiene-d$_{22}$\label{lis:dPSplusdTPB},
\item the same mixture as \ref{lis:dPSplusdTPB}, but annealed at
  75$^\circ$C for 12 hours.
\end{enumerate}
The three WLS films were prepared by dissolving the mixture of the two plastics into toluene at a ratio of 1~g of the plastic mixture to 20~g of toluene, and then dipping one face of an acrylic disc (2.5 cm diameter, 6 mm thick Solacryl SUVT acrylic, manufactured by Spartech Polycast\cite{SolacrylDatasheet}) into the toluene solution.  The thickness of other WLS films prepared in this manner was measured with a spectral reflectometer and found to be in the range of $0.5 - 0.8 \; \mu$m.  References \cite{McKinsey1997} and \cite{Gehman2011} found no significant impact of film thickness in this range on fluorescence spectrum or efficiency.  The protonated PS had an average molecular weight of 280,000, and the deuterated PS had an average molecular weight of 170,000.  In the case of the deuterated films, deuterated toluene-d$_8$ was used to dissolve the plastics.  The deuterated plastics were obtained at 98\% deuteration from Polymer Source Inc.\ (PS) and ISOTEC Stable Isotopes (TPB).  Visible fluorescence spectra were measured with input VUV light at: 125 nm, 160 nm, and 175, nm.  160 nm corresponds to a bright peak in our light source, while 125 and 175 nm correspond to the wavelength of the scintillation light from argon and xenon (two increasingly common materials for the active component of radiation detectors).  The total efficiency curve for converting VUV to visible light was measured from 120 nm to 220 nm.  We performed both of these measurements using the same experimental hardware used in Reference \cite{Gehman2011}.  In short, the apparatus consists of a deuterium lamp, coupled to a holographic grating monochromator.  Light from the output of the monochromator impinged on our three WLS film samples to create visible fluorescence light.  This fluorescence light was then detected with either a spectrometer (to measure the shape of the fluorescence spectrum) or a calibrated photodiode cell (to measure the absolute spectral intensity).  The entire optical train is contained in the same vacuum volume, with pressure between $5 \times 10^{-5}$ and $8 \times 10^{-5}$ Torr at all times during data taking.  The total efficiency curve for each WLS film was calculated under the assumption of a Lambertian angular emission spectrum.  The short wavelength bound for measurement of the total efficiency curve was determined by the transmittance of the magnesium fluoride (MgF$_{2}$) window of the deuterium lamp.  The long wavelength cutoff for that measurement was determined by the appearance of a second order peak from the monochromator grating at half the selected wavelength (we will add a set of order sorting filters to the lamp to remove these second order peaks and extend our future studies to longer wavelengths).

Experimental uncertainties were calculated for each point in the total efficiency curve accounting for:
\begin{enumerate}
\item noise in the measurement of the photodiode current,
\item uncertainty in the calibrated response of the photodiode,
\item spread in the wavelength of the light from the monochromator.
\end{enumerate}
All other sources of uncertainty (including those from the survey of the instrument used to calculate the geometric efficiency of both photon sensors) were small compared to the three listed above and were therefore neglected.  For more detail on the experimental hardware and data analysis, please refer to Reference \cite{Gehman2011}.

\section{Results and Discussion}\label{sec:Results}
\begin{figure}[h]
\begin{center}
\includegraphics[width=\textwidth]{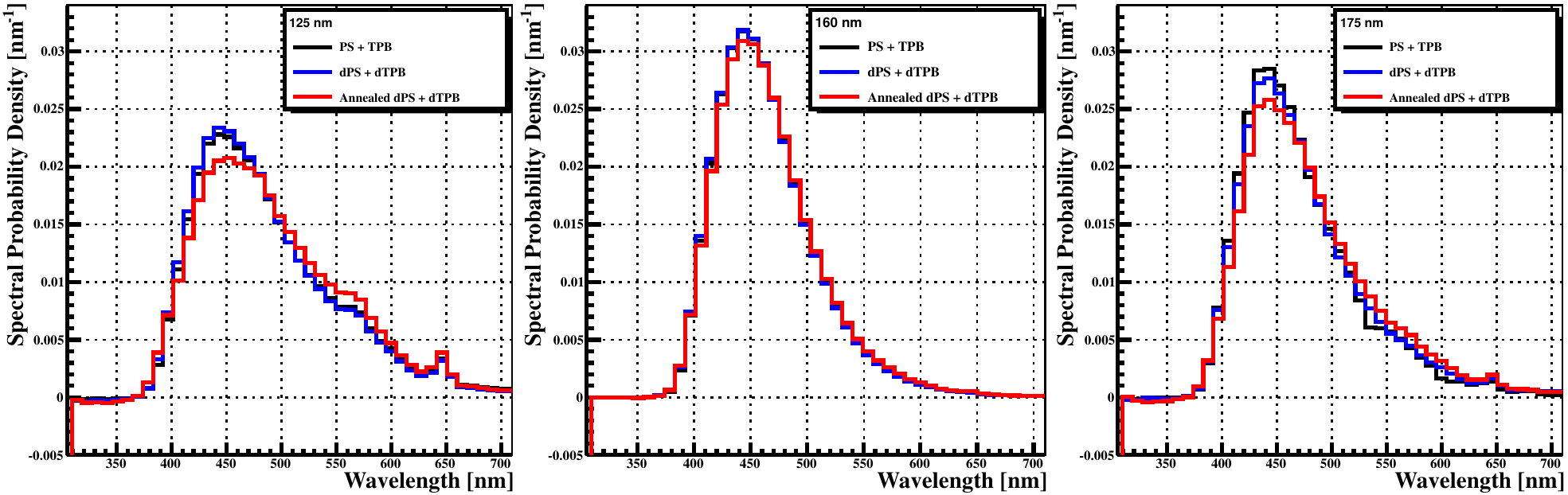}
\caption{Visible fluorescence spectra for all three WLS films at 125 nm (left), 160 nm (middle) and 175 nm (right).}
\label{fig:VisSpec}
\end{center}
\end{figure}

\begin{figure}[h]
\begin{center}
\includegraphics[width=10cm]{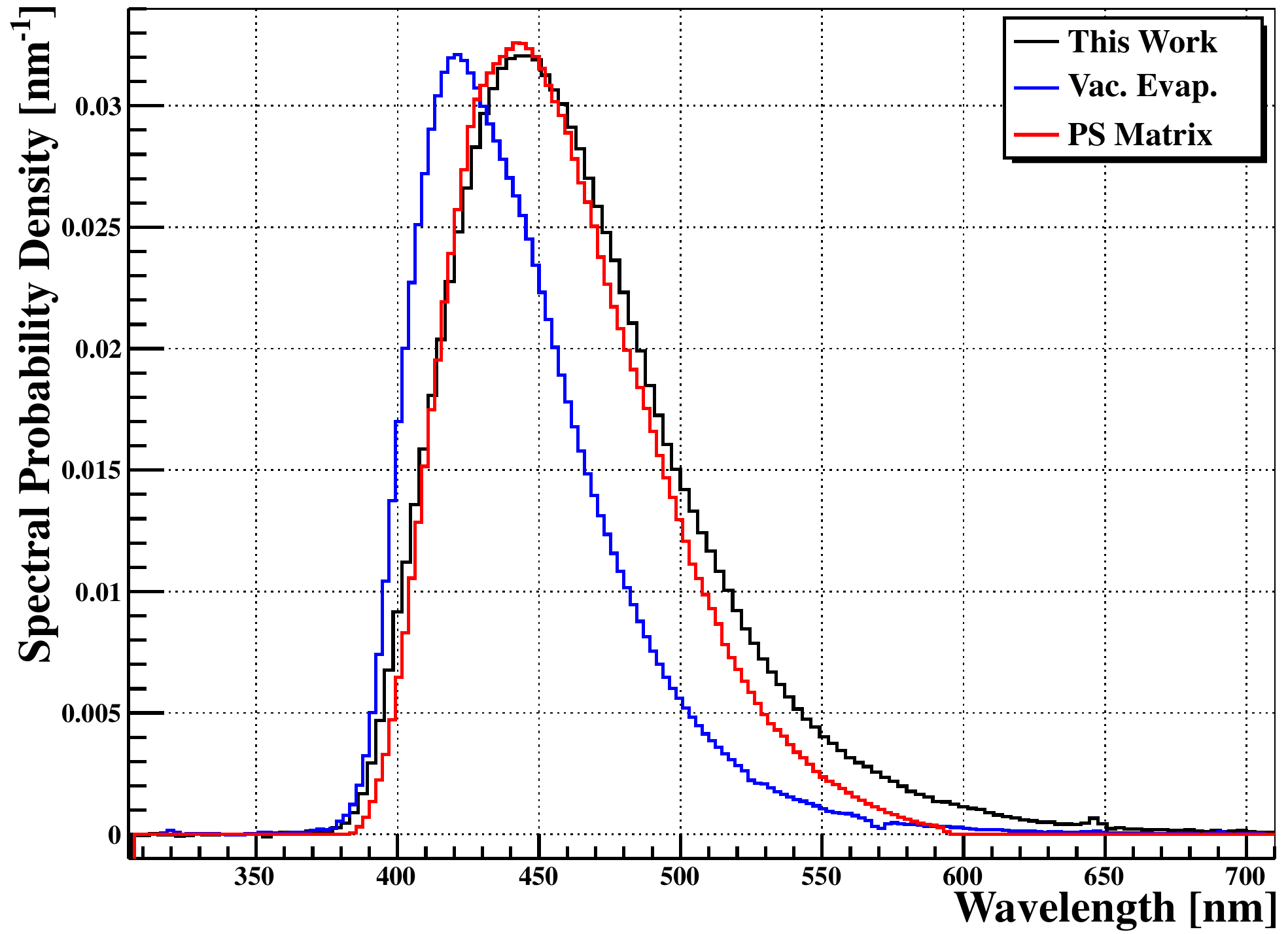}
\caption{Visible fluorescence spectra reproduced from Figure \ref{fig:VisSpec} (labeled ``This Work''), Reference \cite{Gehman2011} (labeled, ``Vac. Evap.''), and Reference \cite{Jones2012} (labeled, ``PS Matrix.'').}
\label{fig:PreviousFlrSpecPlots}
\end{center}
\end{figure}

We plot all nine visible fluorescence spectra (three WLS films at three input wavelengths) in Figure \ref{fig:VisSpec}.
The reader should first note that the fluorescence spectra in Figure \ref{fig:VisSpec} peak at a wavelength consistent with those in Reference \cite{Jones2012} (just less than 450 nm) than those examined in Reference \cite{Gehman2011} (which peak at approximately 420 nm). This clearly follows from the fact that the films in this work were prepared in a similar manner to those in Reference \cite{Jones2012}, as opposed to Reference \cite{Gehman2011}, in which the TPB films were created by vacuum deposition.  The mixture of PS and TPB makes the films far more durable and therefore easy to handle compared to vacuum deposited films.  In addition, for the nEDM experiment, it is necessary to use WLS film made by mixing PS and TPB because the UCN storage cell wall will also act as light guide so the wall need to have optically smooth surfaces.  It appears that the emission spectrum is affected by the environment in which the TPB molecules are placed.  Similar effects have been observed in previous studies in the literature\cite{ElBayoumi1968}.  For convenient comparison by the reader, we reproduce those previous fluorescence spectra here in Figure \ref{fig:PreviousFlrSpecPlots}.

\begin{figure}[h]
\begin{center}
\includegraphics[width=10cm]{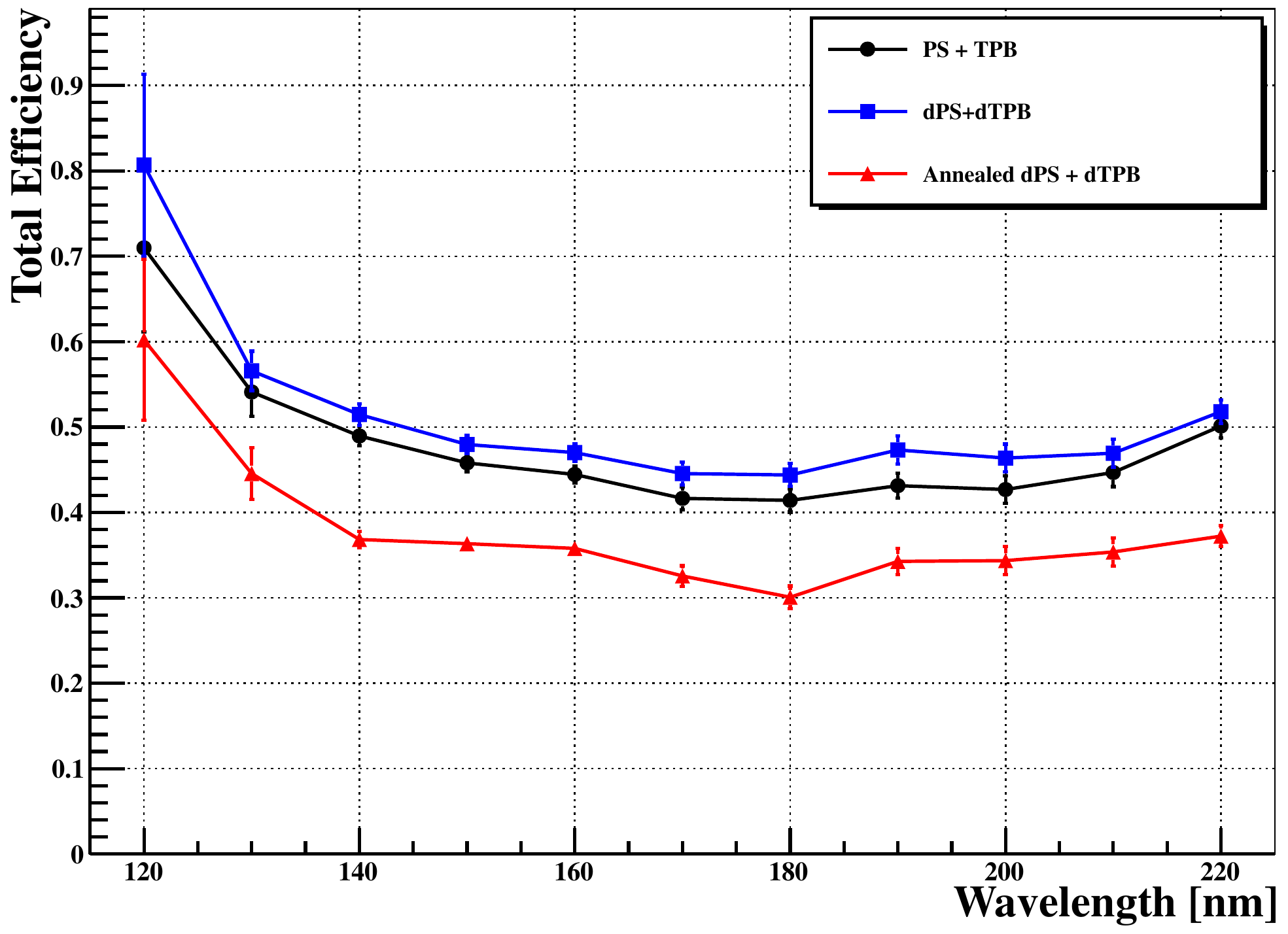}
\caption{Efficiency for converting VUV to visible light for all three WLS films examined in this study.}
\label{fig:TotEff}
\end{center}
\end{figure}

At first glance, the fluorescence spectra have some dependence on the wavelength of input VUV light.  There are features at longer wavelength (a shoulder at $\approx 550$ nm, and a peak at $\approx 650$ nm) present in the 125 and 175 nm spectra that are missing from the 160 nm spectrum.  The intensity of these features seems to scale not with input photon wavelength, but rather with the lamp intensity at those VUV wavelengths.  They also are strongest in the annealed films, which as we will discuss later in the article, have a lower efficiency for converting VUV to visible light and therefore produce dimmer fluorescence spectra.  Recall that the lamp used for this measurement has a strong peak at 160 nm, so the fluorescence spectrum is much more intense than noise from the charged coupled device (CCD) in the spectrometer (because there is a much higher rate of VUV photons incident on the WLS film).  There is a smaller peak in the lamp spectrum at 125 nm, but it is strongly attenuated by the lamp's window (MgF$_{2}$ has a transmittance of $\approx 50$\% at 125 nm).  There is a strong continuum spectrum from the lamp that extends well into the visible band which is unaffected by the MgF$_{2}$ window.  Therefore, the brightest fluorescence spectrum we measure is at 160 nm, followed by 175 nm, with 125 nm being the dimmest.  To collect adequate statistics on the dimmer spectra, we must integrate for longer and sum more CCD exposures.  This allows for transient signals like dark noise or cosmic rays to contaminate the those dimmer spectra more strongly than the brighter one.  For these reasons, we attribute these spectral features near 550 and 650 nm to instrumental artifacts arising from transient backgrounds in a few of the pixels in the CCD.  We therefore use the 160 nm fluorescence spectra for all three WLS films when calculating the total efficiency because it represents, the ``cleanest'' measurement of TPB fluorescence spectrum.  The total conversion efficiency measured for all three WLS films is plotted in Figure \ref{fig:TotEff}.  To our knowledge, this work represents the first stand-alone measurement of the absolute fluorescence efficiency of TPB in a PS matrix. Previous measurements have always measured this efficiency relative to some standard fluor (see e.g. Ref.~\cite{Samson1967}). (For a more detailed discussion on possible uncertainties resulting from the use of a standard fluor, see Ref.~\cite{Gehman2011}).
We first see that the shape of the efficiency curve is very similar to that measured in Reference \cite{Gehman2011}, but the normalization is $\approx 60$\% that measured for vacuum deposited films.

We also note that there is very little difference between the ``standard'' PS + TPB film and the deuterated one.  Finding no significant difference in fluorescence behavior between protonated and deuterated samples is in line with previous work indicating that torsional and vibrational modes involving large structures of the molecule (such as the phenyl rings and the backbone) play an important role in determining the TPB fluorescent efficiency and spectrum\cite{ElBayoumi1968, Rucker1995, Wallace1994, Peng2007}.

On the other hand, it does seem that the process of annealing the film degrades its efficiency to roughly 75\%. We speculate that it is the WLS film, rather than the PMMA substrate, that was affected by the annealing although we do not know what the reason is for the performance deterioration of the WLS film with annealing.

One clear path forward for extensions to this work comes from the fact that helium scintillation light is emitted at 80 nm, well below the short-wavelength cutoff for our lamp in its current configuration.  We are examining the possibility of replacing our current lamp with a windowless, differentially pumped plasma lamp.  Such a light source would not only extend our reach to much shorter wavelengths but also offer greater overall photon intensity, making fluorescence measurements much simpler to collect.

\section{Conclusions}\label{sec:Concl}
We study the effect of deuteration and annealing on the fluorescence spectrum shape and VUV to visible conversion efficiency of TPB films in a polystyrene matrix with input light from 120 to 220 nm.  We observed no discernible difference in the fluorescence spectrum shape between any of the films.  The deuterated film performed equally well compared to the standard one in terms of conversion efficiency, but annealing seems to degrade this efficiency to roughly 75\% of its non-annealed value at all wavelengths studied. 

The problem of collecting the scintillation photons from noble gases can be a rather tricky one because VUV photons are strongly absorbed by most common optical train components.  WLS films are a comparatively simple and inexpensive way to address this problem.  It does, however require a more detailed understanding of their optical propoerties in one's detector.  Measurements like the one described here can help provide that understanding, making possible the next generation of detectors in fundamental physics research.

\section{Acknowledgments}\label{sec:Ack}
This work was supported by Department of Energy under Contract Numbers DE-AC02-05CH11231, 2014LANLEEDM, and DE-AC05-00OR22725.  The authors would like to thank Andrew Hime, Keith Rielage, and Yongchen Sun for their support in the initial design if the experimental apparatus as well as the nEDM publication committee for careful reading and constructive comments.



\begin{thebibliography}{00}
\bibitem{XENON} E. Aprile, {\it et al}. (The XENON100 Collaboration).  Dark Matter Results from 100 Live Days of XENON100 Data.  {\it Physical Review Letters}, {\bf 107} (2011) 131302
\bibitem{XMASS} Jing Liu (for the XMASS collaboration).  The XMASS 800kg Detector.  {\it Journal of Physics: Conference Series}, {\bf 375} (2012) 012022.  DOI:10.1088/1742-6596/375/1/012022
\bibitem{LUX} D. S. Akerib, {\it et al}.  The Large Underground Xenon (LUX) Experiment.  {\it Nuclear Inst. and Methods in Physics Research A}, {\bf 704} (2013) 111.  arXiv:1211.3788 [physics.ins-det].  DOI:10.1016/j.nima.2012.11.135
\bibitem{MiniCLEAN} Jocelyn Monroe.  Recent Progress from the MiniCLEAN Dark Matter Experiment.  {\it Journal of Physisics: Conference Series}, {\bf 375} (2012) 012012.  DOI:10.1088/1742-6596/375/1/012012
\bibitem{DEAP3600} M.G. Boulay (for the DEAP Collaboration).  DEAP-3600 Dark Matter Search at SNOLAB.  {\it Journal of Physics: Conference Series}, {\bf 375} (2012) 012027.  DOI:10.1088/1742-6596/375/1/012027
\bibitem{CLEAN} D.N. McKinsey and K.J. Coakley.  Neutrino Detection with CLEAN.  {\it Astropart. Phys.}, {\bf 22}, 355 (2005).
\bibitem{DARKSIDE} D. Akimov, {\it et al}.  Light Yield in DarkSide-10: a Prototype Two-phase Liquid Argon TPC for Dark Matter Searches.  arXiv:1204.6218 [astro-ph.IM]
\bibitem{ICARUS} ICARUS Collaboration.  A Second-Generation Proton Decay Experiment and Neutrino Observatory at the Gran Sasso Laboratory.  {\it LNGS-P28/2001}, (2001).
\bibitem{MicroBooNE} M. Soderberg, {\it et al} (the MicroBooNE Collaboration).  MicroBooNE: A New Liquid Argon Time Projection Chamber Experiment.  {\it AIP Conference Proceedings}, {\bf 1189} (2009) 83
\bibitem{MEG}Giovanni Signorelli.  The MEG experiment at PSI: status and prospects.  {\it Journal of Physics G}, {\bf 29} (2003) 2027.  DOI:10.1088/0954-3899/29/8/395
\bibitem{EXO} M. Auger, {\it et al}. (EXO Collaboration).  Search for Neutrinoless Double-Beta Decay in $^{136}$Xe with EXO-200.  {\it Physical Review Letters}, {\bf 109} (2012) 032505.  DOI:10.1103/PhysRevLett.109.032505
\bibitem{NEXT} Theopisti Dafni (the NEXT Collaboration).  NEXT: a Gaseous-Xe TPC for the Neutrinoless Double-Beta Decay.  {\it Nuclear Physics B - Proceedings Supplements}.  {\bf 229Ð232} (2012) 479.  DOI:10.1016/j.nuclphysbps.2012.09.116
\bibitem{LBNE} B. Yu, {\it et al}.  Designs of Large Liquid Argon TPCs Ñ from MicroBooNE to LBNE LAr40.  {\it Physics Procedia}, {\bf 37} (2012) 1279.  DOI:10.1016/j.phpro.2012.03.737
\bibitem{Lanou1987} R.E. Lanou, H.J. Maris, and G.M. Seidel.  Detection of Solar Neutrinos in Superfluid Helium.  {\it Physical Review Letters}, {\bf 58} (1987) 2498
\bibitem{McKinsey2000} D.N. McKinsey and J.M. Doyle.  Liquid Helium and Liquid Neon---Sensitive, Low Background Scintillation Media for the Detection of Low Energy Neutrinos.  {\it Journal of Low Temperature Physics}, {\bf 118} (2000) 153

\bibitem{nEDM1} T.M. Ito (for the nEdm Collaboration).  Plans for a Neutron EDM Experiment at SNS.  {\it Journal of Physics: Conference Series}, {\bf 69} (2007) 012037.  DOI:10.1088/1742-6596/69/1/012037.  http://p25ext.lanl.gov/edm/edm.html
\bibitem{nEDM2} R. Golub and S.K. Lamoreaux.  Neutron Electric Dipole Moment, Ultracold Neutrons and Polarized 3He.  {\it Physics Reports}, {\bf 237}(1994) 1
\bibitem{Huffman2000} P.R. Huffman, {\it et al}.  Magnetic Trapping of Neutrons.  {\it Nature}, {\bf 403} (2000) 62
\bibitem{Lippincott2010} W.H. Lippincott.  Direct Detection of Dark Matter With Liquid Argon and Neon.  PhD Thesis.  Yale University (2010)
\bibitem{McKinsey1997} D.N. McKinsey, C.R. Brome, J.S. Butterworth, R. Golub, K. Habicht, P.R. Huffman, S.K. Lamoreaux, C.E.H. Mattoni, J.M. Doyle.  Fluorescence Efficiencies of Thin Scintillating Films in the Extreme Ultraviolet Spectral Region.  {\it Nuclear Instruments and Methods in Physics Research B}, {\bf 132} (1997) 351
\bibitem{Jones2012} B.J.P. Jones, J.K. VanGemert, J.M. Conrad, A. Pla-Dalmau.  Photodegradation Mechanisms of Tetraphenyl Butadiene Coatings for Liquid Argon Detectors.  Submitted to {\it Journal of Instrumentation}.  arXiv:12117150 [physics.ins-det]
\bibitem{Gehman2011} V.M. Gehman, S.R. Seibert, K. Rielage, A. Hime, Y. Sun, D.-M. Mei, J. Maassen, and D. Moore.  Fluorescence Efficiency and Visible Re-emission Spectrum of Tetraphenyl Butadiene Films at Extreme Ultraviolet Wavelengths.  {\it Nuclear Instruments and Methods in Physics Research A}, {\bf 654} (2011) 116.  DOI:10.1016/j.nima.2011.06.088.  arXiv:1104.3259 [astro-ph.IM]
\bibitem{McKinsey2004} D. N. McKinsey, C. R. Brome, J. S. Butterworth,
  S. N. Dzhosyuk, R. Golub, K. Habicht, P. R. Huffman,
  C. E. H. Mattoni, L. Yang, J. M. Doyle. Detecting ionizing radiation
  in liquid helium using wavelength shifting light collection. {\it Nuclear Instruments and
  Methods in Physics Research A}, {\bf 516} (2004) 475.
\bibitem{Archibald2006} G. Archibald, J. Boissevain, R. Golub,
  C. R. Gould, M. E. Hayden, E. Korobkina, W. S. Wilburn,
  J. Zou. Characterizing of Scintillation Light Produced in Superfluid
  Helium-4. AIP Conference Proceedings {\bf 850} (2006) 143. 
\bibitem{Ito2012} T. M. Ito, S. M. Clayton, J. Ramsey, M. Karcz,
  C.-Y. Liu, J. C. Long, T. G. Reddy, G. Seidel. Effect of an electric
  field on superfluid helium scintillation produced by
  $\alpha$-particle sources. {\it Physical Review A} {\bf
    85} (2012) 042718.
\bibitem{Golub1991} R.Golub, D. Richardson, S.K.Lamoreaux.  Ultra-Cold Neutrons.  Adam Hilger, Bristol (1991).
\bibitem{Ignatovich1990} V.K. Ignatovich.  The Physics of Ultracold Neutrons.  Clarendon Press, Oxford (1990).
\bibitem{Lamoreaux1988} S.~K.~Lamoreaux, Institut Laue Langevin Report
  ILL88LA1T, 1988.
\bibitem{SolacrylDatasheet}Spartech Polycast.  Solacryl Stabilized UV Transmitting Acrylic Datasheet. http://www.spartech.com/polycast/Spartech-Polycast-SUVT.pdf
\bibitem{Samson1967} J. A. R. Samon, Techniques of Vacuum Ultraviolet
  Spectroscopy, John Wiley \& Sons Inc., 1967.
\bibitem{ElBayoumi1968} M.A. El Bayoumi and F.M.A. Halim.  Effect of Intramolecular Twisting on the Fluorescence Spectra of Sterically Hindered Tetraphenylbutadienes.  {\it Journal of Chemical Physics}, {\bf 48} (1968) 2536
\bibitem{Rucker1995} R.L. Rucker, B.J. Schwartz, M.A. El Bayoumi, C.B. Harris.  Ultrafast Dynamics of Sterically Hindered Tetraphenylmethylbutadiene in Liquids.  {\it Chemical Physics Letters}, {\bf 235} (1995) 471
\bibitem{Wallace1994} S.E. Wallace-Williams,B.J. Schwartz, S. M\/{o}rller, R.A. Goldbeck, W.A. Yee, M.A. El-Bayoumi, and D.S. Kliger.  Excited State Spectra and Dynamics of Phenyl-Substituted Butadienes.  {\it Journal of Physical Chemistry}, {\bf 98} (1994) 60
\bibitem{Peng2007} Q. Peng, Y. Yi, Z. Shuai, and J. Shao.  Toward Quantitative Prediction of Molecular Fluorescence Quantum Efficiency: Role of Duschinsky Rotation.  {\it Journal of the American Chemical Society}, {\bf 129} (2007) 9333
\end{thebibliography}
\end{document}